\documentclass[	a4paper]{article}
\usepackage[english]{babel}
\usepackage{icrc2013}

\title{UHECR correlations taking account of  composition and Galactic magnetic deflections}

\shorttitle{UHECR and Galactic magnetic deflections}

\authors{
J. Roberts$^{1,2}$ (presenter) and G. Farrar$^{2}$
}

\afiliations{
$^1$ New York University, Abu Dhabi;  ~
$^2$ Center for Cosmology and Particle Physics, New York University, NY, NY
}

\email{jonathan.roberts@nyu.edu}

\abstract{We predict the arrival direction distribution of cosmic rays including their deflection in the Galactic magnetic field, for several combinations of source and composition hypotheses:  the sources are hard X-ray AGNs or uniformly sample the matter distribution of galaxies, and the composition at the source is pure proton or is the Galactic cosmic ray composition measured by CREAM.   We use the regular component of the Jansson-Farrar 2012 model for the GMF and allow for rigidities as low as 2 EeV.   We report the correlations of published UHECRs, rescaling event energies so as to reconcile the spectra of the different experiments and taking the overall energy uncertainty into account; different composition hypotheses are considered. This work demonstrates the feasibility of calculating GMF deflections to low enough rigidities to allow for heavy composition in correlation studies, and that non-trivial arrival direction structure should be expected even for mixed or heavy composition, as long as UHECRs come from the local universe.}


\begin{document}
\maketitle

\section{Introduction}

The predicted arrival directions of Ultrahigh Energy Cosmic Rays (UHECRs) are sensitive to assumptions on the distribution of sources, composition at the source, propagation to our galaxy, and deflection in the magnetic field of the Galaxy. We present a correlation analysis framework that allows for all of the above, and use it to predict the sky map of the cosmic ray distribution for two different models of the source distribution.  Finally, we compare with published UHECR data.

\section{Source Distribution}

\begin{figure}[!h]
  \centering
  \includegraphics[width=0.45\textwidth]{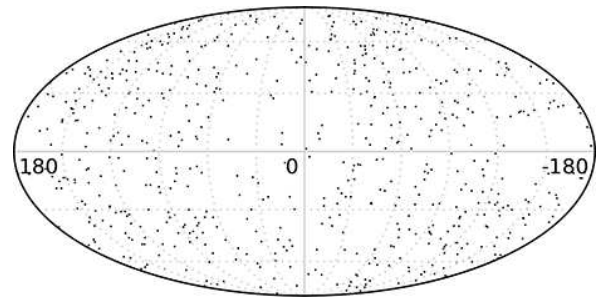}
  \caption{Distribution of extragalactic hard X-ray AGN candidate sources from the Swift-BAT catalog.}
  \label{BAT}
 \end{figure}

One hypothesis for the source distribution is that the cosmic rays are produced by hard X-ray AGNs. To test this we use the Swift-BAT catalog of hard X-ray sources \cite{Baumgartner:2012qx}, and correct the distances for peculiar motion. From this we make a sub-catalog of the Swift-BAT sources that is volume limited out to $z=0.018$ [or 76 Mpc], and includes all sources beyond 76 Mpc; a total of 574 objects. A skymap of these candidate sources is shown in Fig.~\ref{BAT}. The sources are sparse, which makes the catalog highly predictive. However, if the catalog is incomplete, or if we have guessed wrongly about the subset of galaxies that produce cosmic rays, then a correlation study will not find a good match. For the purpose of this study we take UHECR production to be uniform across all sources in flux, spectrum and composition.

\begin{figure}[!h]
  \centering
  \includegraphics[width=0.45\textwidth]{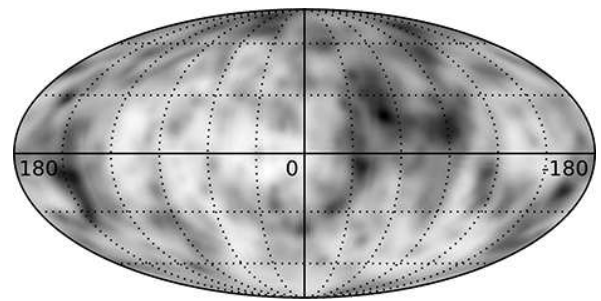}
  \caption{Galaxy density from the Willick density map, integrated along the lines of sight out to 120 Mpc/h.}
  \label{sources}
 \end{figure}

The second hypothesis we consider is that UHECRs are produced by unspecified sources that track the overall galaxy density. We take the density field of galaxies from two density maps created from the IRAS 1.2 Jy redshift survey \cite{Willick:1996km} and from the 2MRS \cite{Masters}.  Each of these master density maps provides a a three dimensional binned density of galaxies in our neighborhood. The closer map, created by Willick et al. \cite{Willick:1996km}, has a higher resolution, and extends out to 120Mpc/h. Beyond 120Mpc/h we use the catalog created by Masters et al. \cite{Masters}, which extends to 200Mpc/h. Beyond 200Mpc/h the density of galaxies is taken to be homogeneous and isotropic. Fig~\ref{sources} shows the density integrated along the line of sight out to 120Mpc/h.  Both Figs. 1 and 2 are in galactic coordinates.

\subsection{A Note on Peculiar Motion}

The peculiar motion of galaxies presents a three-fold problem when constructing a predicted skymap of UHECRs.  At large distances, the redshift is predominantly cosmological and peculiar motion can be ignored and the Hubble relation gives a good estimate of distance. For nearby galaxies, however, the redshift is often dominated by peculiar motion; if peculiar motion is ignored and the redshift alone is used to estimate distance, the distance of sources can be mis-estimated by a factor of 10 or more.  Since for a given luminosity in UHECRs the flux at Earth $\sim 1/r^2$, this can produce factor 100 errors in the predicted CR flux from a source.  In addition, an inaccurate distance determination can affect whether a source appears in the volume-limited catalog.  Finally, at high energies where propagation losses are important, the probability of a UHECR arriving with its observed energy from a given source is strongly dependent on the source distance.

For the Swift-BAT AGNs we find distances on a case by case basis, from determinations in the literature reported in NED when available, or when they can be associated with a galaxy cluster we use the cluster distance. The density maps of Masters et al. and Willick et al. were created by a self-consistent procedure of computing the three dimensional map of peculiar velocities for a given density assumption, and thus they account for the effect of peculiar motion. 

\section{Composition and Propagation}

We create the local spectrum of cosmic rays by propagating 100,000 cosmic rays in steps of 2.5~Mpc. We use CRPropa \cite{Kampert:2012fi} to model the energy loss as UHECRs propagate from source to Earth. We take the injection spectrum of cosmic rays to be proportional to $E^{-2.3}$, and consider two models for the composition of the cosmic rays at the source - either pure proton, or a ratio of charges derived from the Galactic ratios of cosmic rays measured by the CREAM experiment \cite{CREAM}. Table \ref{CREAM_ratios} gives the relative abundance of each charge at the source.

\begin{table}[!h]
\centering
\begin{tabular}{c|c}
Element & Relative Abundance\\
\hline
H	&	5\\
He	&	5\\
C	&	0.429\\
N	&	0.038\\
O	&	1\\
Ne	&	0.208\\
Mg	&	0.448\\
Si	&	0.793\\
Fe	&	3.844\\
\end{tabular}
\caption{The relative abundance (normalized to oxygen) of galactic cosmic rays measured by the CREAM experiment, hereafter referred to as the CREAM model.}
\label{CREAM_ratios}
\end{table}

For the BAT model, we propagate cosmic rays from each source.  For the generic galaxy model we sum the spectrum along all distance steps weighted by the galactic density map, for each pixel in the resolution 9 Healpix sky map. 

\begin{figure}[!h]
  \centering
  \includegraphics[width=0.45\textwidth]{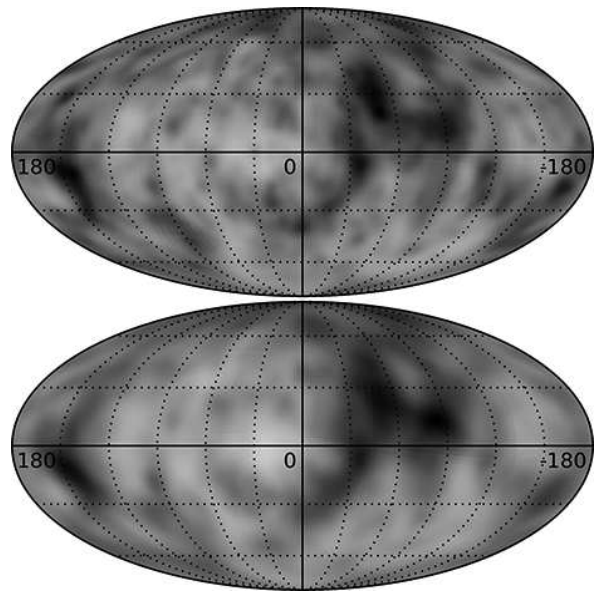}
  \caption{The simulated distribution of undeflected cosmic rays with $Z=1$ (top) and $Z=20-26$ (bottom) that reach the Galaxy with energy in the range $47.3<E(\mathrm{EeV})<53.1$, for the ``CREAM'' model of particle injection at the source.}
  \label{undeflected_allsky}
 \end{figure}

The energy loss of UHECRs is highly charge-dependent. In Fig.~\ref{undeflected_allsky} we plot the density of protons (above) and heavy UHECRs (Z=20-26, below) that reach our galaxy in the energy range $47.3<E(\mathrm{EeV})<53.1$. The summed all-sky flux of heavy CRs with energy $47.3<E(\mathrm{EeV})<53.1$ is approximately a quarter of the all-sky proton flux.  The simulated proton skymap displays finer structures than the map of heavy composition. This is because proton UHECRs in this energy range have longer path lengths, and can originate in more distant structures, which consequently have a smaller angular size on the sky.

\begin{figure}[!h]
  \centering
  \includegraphics[width=0.45\textwidth]{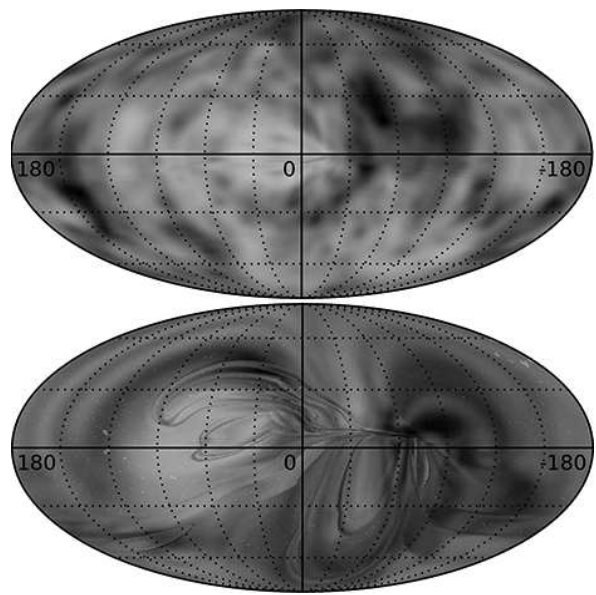}
  \caption{The distribution of cosmic rays shown in Fig.~\ref{undeflected_allsky}, after deflections in the JF12 Galactic magnetic field.}
  \label{deflected_allsky}
 \end{figure}

\section{Calculating Galactic Magnetic Deflections}

We model the deflections of the cosmic rays using the Jansson-Farrar model for the regular component of the Galactic magnetic field \cite{GMF}.  \footnote{Smearing due to random fields is expected to be small compared to the total deflection in the coherent component \cite{TTrand}.} The forward tracking is calculated by backtracking a resolution 11 Healpix grid of cosmic rays at evenly spaced steps in $\log_{10}E$. Since the energy losses of UHECRs in the Galaxy are negligible, Liouville's theorem applies. We take advantage of Liouville's theorem to convert the backtracking map into a forwardtracking map, guaranteeing that an isotropic distribution of cosmic rays incident on the galaxy will produce an isotropic distribution of incident cosmic rays at Earth \cite{FJR13}.  

Solving the forwardtracking problem enables the precise prediction of magnetic deflections for UHECRs from any source, and the calculation of magnification and multiple imaging caused by the Galactic magnetic field.  We utilize the forwardtracking calculation to generate an all-sky map of the predicted flux of UHECRs, for each energy step and charge. Fig.~\ref{deflected_allsky} shows the forward-tracked all-sky maps for proton cosmic rays (top) and heavy cosmic rays (Z=20-26, bottom) in the energy band $47.3<E(\mathrm{EeV})<53.1$.

Comparing Figs.~\ref{undeflected_allsky} and \ref{deflected_allsky} shows that propagation through the GMF distorts the skymaps and results in distinct structures, even for the high charge component. Evidently, the common assumption that iron UHECRs have an effectively isotropic distribution is overly simplistic, at least in the approximation of neglecting the random component of the GMF.
 
 \begin{figure}[!h]
  \centering
  \includegraphics[width=0.45\textwidth]{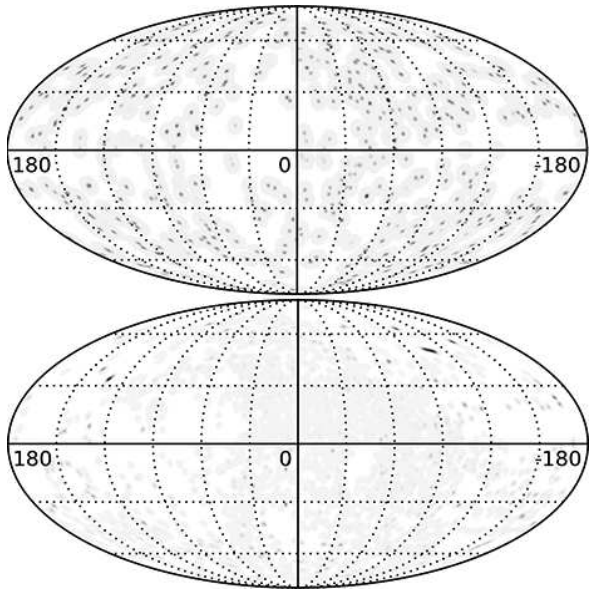}
  \caption{Distribution of protons (top) and heavy ($Z=20-26$) cosmic rays (bottom) in the energy range $47.3<E(\mathrm{EeV})<53.1$, with the BAT distribution of sources and CREAM composition, after deflection in the GMF. In each plot, black designates the highest density of simulated UHECRs. The two plots are separately normalized.}
  \label{BAT_distribution}
 \end{figure}
 
The BAT source hypothesis shows a similar story. In Fig.~\ref{BAT_distribution} we plot the distribution of protons and heavy UHECRs in the energy range $47.3<E(\mathrm{EeV})<53.1$. Here the cosmic ray arrival directions have been smeared by a 2D Gaussian in which 68\% of the distribution is within 1 degree of the center. This makes the individual sources easier to plot, and models a 1 degree uncertainty in the measurement of the arrival direction. The distribution of UHE protons follows the BAT sources relatively closely - it's easy to make out the super-galactic plane for example. The distribution of deflected heavy UHECRs is smoother, but is far from isotropic.

\section{Correlations}

\subsection{Analysis}

We take the published events from the Auger (69 events) \cite{Abreu:2010ab}, Telescope Array (17 events) \cite{AbuZayyad:2012hv}, HiRes (10 events) \cite{HiRes} and AGASA (10 events) \cite{Hayashida:2000zr} collaborations, with energies corrected according to the recommendation of the Spectrum Working Group \cite{EnergyCorrections}. We take the angular uncertainty for Auger, Telescope Array and HiRes events to be 1 degree, and for AGASA to be 1.5 degrees, modeled as a 2D Gaussian.  Energy resolution is described by
\begin{equation}
q(E_0,E)=\frac{1}{\lambda\sqrt{2\pi}}\exp\left(-\frac{(\epsilon-\epsilon_0)^2}{2\lambda^2}\right),
\end{equation}
where $E$ and $E_{0}$ are respectively the true and observed energy, $\epsilon=\ln(E)$ and $\lambda$ is the fractional energy uncertainty.

For each cosmic ray we define the quantity
\begin{equation}
p(E_0,\mathbf{r}_0)=\int \sum_k P(E,\mathbf{r}) q(E_0,E_k) Q(\mathbf{r}_0,\mathbf{r}) d^2\mathbf{r},
\end{equation}
where $k$ labels the energy bins (of width $\Delta \log_{10} E = 0.05$), $\mathbf{r}_0$ is the observed direction of the cosmic ray, $Q(\mathbf{r}_0,\mathbf{r})$ is a 2D Gaussian that models the measurement uncertainty in the direction of the cosmic ray, $P(E_{k},\mathbf{r})$ is a set of values from the deflected sky maps that correspond to the likelihood of seeing a cosmic ray with an energy in bin $k$ from direction $\mathbf{r}$, including forwardtracking through the GMF, extragalactic energy losses and photodisintegration. $P$ is derived from the number density of simulated CRs that arrive in each pixel within each energy bin, which provides the relative normalization of $p$ across all arrival directions and all energy steps.

For each cosmic ray $i$ we calculate $p_{i}$, under our different hypotheses on sources and composition and also assuming an isotropic distribution, $p_{iso}$. The ratio of the different $p$'s provides a measure of the relative likelihood of each hypothesis, for each individual cosmic ray.   We can also consider the information in the ensemble of events, by defining the likelihood measure
\begin{equation}
LM=\sum_i w \, ln\, p_{src,i} + (1-w) \, ln \, p_{iso,i}.
\end{equation}
Here, $w$ is the fraction of events coming from the source distribution for some source and composition ansatz, compared to an isotropic distribution; $w$ is calculated by maximizing the likelihood measure $LM$.  More details of the method, in particular the normalization procedures, treatment of exposure, etc, can be found in \cite{GML}.

\subsection{Sources vs Isotropic}

\begin{figure}[!h]
  \centering
  \includegraphics[width=0.45\textwidth]{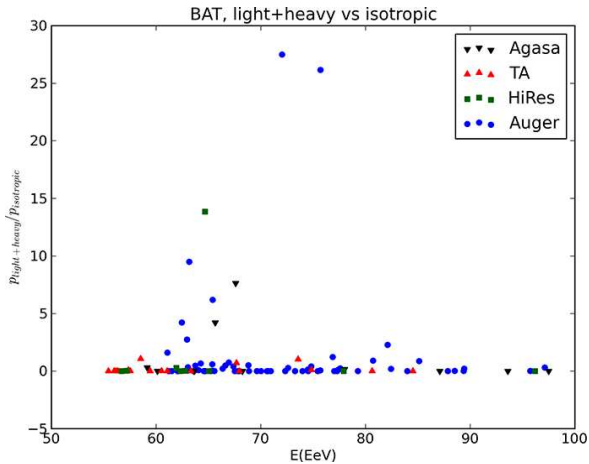}
   \caption{The relative value of $p_{iso}$ vs $p_{src}$ for each cosmic ray in the sample, assuming sources are BAT hard X-Rays and the cosmic rays are produced at source with an CREAM ratio of elemental compositions. \label{BAT_all_vs_iso}}
\end{figure}

Fig.~\ref{BAT_all_vs_iso} shows the ratio $p_{src}/p_{iso}$ for each cosmic ray in the combined data set, assuming sources are BAT hard X-Rays and the cosmic rays are produced with a CREAM ratio of elemental compositions.  The majority of values are close to zero, due to the sparseness of the set of BAT hard X-ray sources and the assumed small $1^{\circ}$ directional uncertainty.  Maximizing the likelihood function gives a value for the source fraction of 2.5\%. In comparison, maximizing the likelihood for each of 100 sets of mock UHECRs in which sources are BAT hard X-Rays producing a CREAM composition ratio results in a correlation fraction of 56\% or above. Allowance for extragalactic angular deflections and the systematic uncertainty on deflections due to imperfect knowledge of the GMF could be included by scanning over the smearing angle but we do not do so here. 

\begin{figure}[!h]
  \centering
  \includegraphics[width=0.45\textwidth]{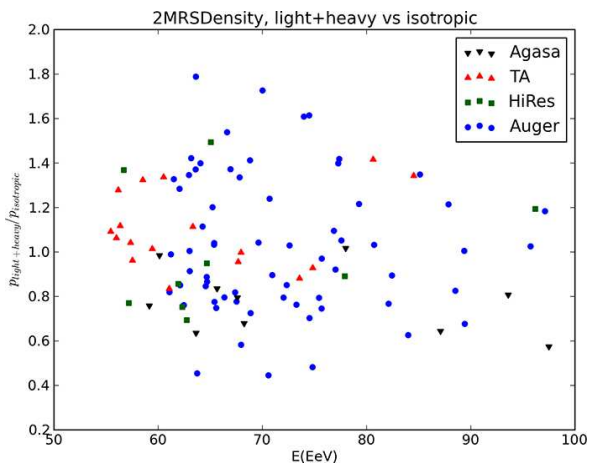}
   \caption{The relative value of $p_{src}$ and $p_{proton}$ for each cosmic ray in the sample, assuming a 2MRS density distribution of sources.\label{2MRS_all_vs_iso}}
\end{figure}

Fig.~\ref{2MRS_all_vs_iso} shows the relative values for $p_{source}/p_{iso}$ taking sources to be distributed like the 2MRS density distribution of galaxies, also for the CREAM composition hypothesis. The data shows a smooth continuum, with some events more likely to be from the 2MRS density with a CREAM composition ($p_{src}/p_{iso}>1$) and others better described by an isotropic distribution ($p_{src}/p_{iso}<1$). Maximizing the likelihood function gives a value for the source fraction of 19\%. In comparison, maximizing the likelihood function for mock UHECRs drawn from the same hypothesis gives a source fraction of 100\% in all 100 mock datasets.

\section{Summary and future work}

We have presented first results of a program to study UHECR sources and composition, taking into account deflections in the galactic magnetic field and peculiar velocity corrections for the sources, and considering several types of source, spectrum, and composition hypotheses. The simplest case has been completed, considering only the regular component of the galactic magnetic field, ignoring extra-galactic deflections and taking either a pure proton spectrum or a mixed composition with relative abundances corresponding to the CREAM measurement of relativistic Galactic cosmic rays. In future work we will extend this to consider a larger range of hypotheses on source spectrum and composition, extra-galactic deflections, and the effect of a random component of the Galactic magnetic field.  

This work shows the feasibility of calculating GMF deflections to low enough rigidities to allow for heavy composition in correlation studies.  It also shows that non-trivial arrival direction structure is expected even for mixed or heavy composition, as long as UHECRs come from the local universe as expected for both protons and nuclei due to photoproduction and photodisintegration energy losses.   Thus when the number of UHECRs becomes large enough, UHECR anisotropy should be found independently of the composition.  If no anisotropy is found, it will suggest that the UHECR spectral cutoff is due to a cutoff in accelerating power of sources rather than energy losses during propagation.

\vspace*{0.5cm}
\footnotesize{{\bf Acknowledgment:}{This work was supported by the grants NASA NNX10AC96G, NSF PHY-1212538, NSF PHY-0900631 and NSF PHY-0970075.  We are grateful to M. Sutherland and A. Keivani for help in calculating GMF deflections. }}

\end{document}